\documentstyle[twocolumn,epsf]{jpsj}
\def\runtitle{ 
Electronic states in correlated three-coupled chains
}
\def\runauthor{
M. Tsuchiizu, Y. Suzumura
}

\def\td{t_{\rm d}}
\def\tperp{t_{\perp}}

\def\e{{\rm e}}

\def\virg{\;\;,}

\def\vf{v_{\rm F}}
\def\kf{k_{\rm F}}

\def\ggs{\buildrel\textstyle > \over {\hbox{\raise0.2ex\hbox{$\sim$}}}}
\def\lls{\buildrel\textstyle < \over {\hbox{\raise0.2ex\hbox{$\sim$}}}}
\def\gsim{\,\lower0.75ex\hbox{$\ggs$}\,}
\def\lsim{\,\lower0.75ex\hbox{$\lls$}\,}
\def\para{\parallel}

\def\ttperp{\tilde{t}_\perp}

\def\Gbppzz {G_{1C}}

\def\Gbpzpm {G_{1H}}

\def\Gfpzzp {G_{2E}}

\def\Gfpzpm {G_{2H}}

\def\Gppzzp {G_{\para E}}

\def\Gppzpm {G_{\para H}}
\def\Gupppp {G_{3A}}
\def\Guzzzz {G_{3B}}
\def\Guppzz {G_{3C}}
\def\Guppmm {G_{3D}}
\def\Gupzzp {G_{3E}}
\def\Gupmmp {G_{3F}}
\def\Guzpzm {G_{3G}}
\def\Gupzpm {G_{3H}}
\def\cite #1{[\citen{#1}]}

\def\jo #1#2#3#4{#1 {\bf #2} (#3) #4} 

\def\PRB{Phys.\ Rev.\ B}

\def\JPSJ{J.\ Phys.\ Soc.\ Jpn.}

\def\PTP{Prog.\ Theor.\ Phys.}

\def\PLA{Phys.\ Lett.\ A}

\def\PSS{Phys.\ Status\ Solidi\ B}

\title
{
Electronic states in correlated three-coupled chains
 }

\author{
M. Tsuchiizu$^a$, Y. Suzumura$^{a,b}$
}

\inst{
$^{a}$Department of Physics, Nagoya University, Nagoya 464-8602, Japan
\\
$^{b}$CREST, Japan Science and Technology Corporation (JST) 
}

\recdate{\hspace{3cm}}

\abst{
Three-coupled chains of 
  quarter-filled Hubbard model with dimerization have  been 
    studied for periodic boundary conditions. 
By use of a renormalization group method, it is found that 
 the interchain hopping is renormalized to zero leading to 
  confinement when charge gap becomes larger than the effective 
 interchain   hopping energy.
Such a result  of confinement is compared with that of two-coupled 
  chains.
 }

\kword{
A. organic compounds, D. electronic structure, D. phase transition
}

\begin{document}
\maketitle

  Spin gap and charge gap have  been studied extensively 
    in coupled-chain  systems. 
Theoretical study  in terms of  
 a  renormalization group (RG) method shows   
 the spin gap in two-coupled chains 
   for small but relevant interchain hopping \cite{KR}
   where the electronic state is given by 
     $d_{x^2-y^2}$-like superconducting state.
 The spin gap has been  also examined 
 for three-coupled chains \cite{Arrigoni,Schulz2,Kimura},
 whose  state depends on boundary conditions 
  in the transverse direction. 
 In case of  small interchain hopping at half-filling, 
 there are both charge gap and spin gap 
  for open boundary conditions (OBC) while  charge gap 
   vanishes  for  periodic boundary conditions (PBC).
 A system with infinite chains at half-filling exhibits  
  charge gap but no spin gap \cite{Bourbonnais_U}.

Although these works treat  the relevant interchain hopping, 
 a theory also suggests 
  the irrelevant one-particle interchain hopping for  half-filling
 \cite{Bourbonnais_U}. 
 In order to understand  optical experiments  on 
     quasi-one-dimensional organic conductors \cite{Vescoli,Schwartz}
   indicating  a  confinement - deconfinement transition,  
      two-coupled chains at half-filling have been studied 
 where  a transition 
  from deconfinement to confinement occurs
 for   umklapp scattering being  larger than a critical value 
 \cite{Suzumura,Tsuchiizu_SCES}.
In the present paper, 
  such a transition is examined 
  for three-coupled chains.
Although the results of two chains seem to explain the experiments,
  it is not yet known if such a transition can be expected for 
  many chains.

The Hamiltonian for quarter-filled three-coupled chains 
 with on-site repulsion ($U$) and interchain hopping ($\tperp$) 
 is written as 
\begin{eqnarray}
{\cal  H} &=& - \sum_{j,\sigma,l}
   \left[ t + (-1)^{j} \td\right] 
   \left(  c_{j \sigma l}^{\dagger} 
    \,\, c_{j+1 \sigma l} + \mbox{\rm h.c.} \right)
\nonumber \\ &&{}
- t_{\perp} \sum_{j,\sigma,l} 
\left(
c_{j \sigma l}^{\dagger} \,\,
 c_{j \sigma l+1}
 + \mbox{\rm h.c.} \right)
+ U \sum_{j,l} \, n_{j \uparrow l} \, n_{j \downarrow l},
\label{eq:H}
\end{eqnarray}
 where 
  $n_{j \sigma l} = c_{j \sigma l}^\dagger c_{j \sigma l}$ and
  $c_{j \sigma l}$ denotes the annihilation operator of the electron
   at the $j$-th site
  of the $l$-th chain ($l=$1, 2, 3) with spin 
  $\sigma$($=\uparrow,\downarrow$), and $c_{j\sigma 4}=c_{j \sigma 1}$.
 In Eq.~(\ref{eq:H}), 
 $\td$ denotes dimerization along the chains and 
   the case for PBC  is studied.

 Diagonalizing the $\td$-term, which leads effectively to 
  the half-filled band \cite{Tsuchiizu_SCES},
 we consider an  effective Hamiltonian ${\cal H}^d$ consisting of 
  the lower band  with   
        fermion operators of  $d_{k \sigma l}$. 
 The terms for the interchain hopping  can be diagonalized
  by introducing 
  Fourier transform, 
$a_{k\sigma\mu} \equiv (1/\sqrt{3})$ $  \sum_{l=1}^{3} 
  \exp\left[-i k_y(\mu) \, l \right] d_{k\sigma l}$
 with  $k_y (\mu) = (2\pi/3) \mu$ $(\mu =0,\pm 1)$.  
 The kinetic term  is written as 
$ {\cal H}_K^d \equiv 
\sum_{k,\sigma,\mu} $ $\varepsilon(k,k_y) $ $   
       a_{k \sigma \mu}^\dagger $ $a_{k \sigma \mu}$
with  
$
\varepsilon(k,k_y)
= -2 \sqrt{t^2 \cos^2 ka + \td^2 \sin^2 ka}-2\tperp \cos k_y
$, 
which is rewritten, in terms of  the  linearized   dispersion,  as  
${\cal H}_K^d = 
\sum_{k,p,\sigma,\mu} $ $\vf (pk-k_{{\rm F}\mu}) \,
  a_{k p \sigma \mu}^\dagger \, a_{k p \sigma \mu}$
  with  $p$ being  the index of the branch $p= +$ $(-)$ corresponding to
  right moving (left moving) electrons.
Fermi momenta are given by
$k_{{\rm F}0} = \kf + 2\tperp/\vf$ and
$k_{{\rm F}\pm} = \kf - \tperp/\vf$ where 
  $\vf=\sqrt{2}ta$ $[1-(\td/t)^2]$ $/\sqrt{1+(\td/t)^2}$
   \cite{Tsuchiizu_SCES}.
Following the conventional $g$-ology, 
 coupling constants are given by
$g_{1^\perp} =  g_{2^\perp} = g_{4^\perp} =  U  a$,
$g_3 = U a ( 2\td/t ) /[1+(\td/t)^2]$ \cite{Tsuchiizu_SCES} and
$g_{1^\para} = g_{2^\para} = g_{4^\para}= 0$.

Applying the bosonization method,
 we introduce phase variables 
   $\theta_{\rho \mu}$ and  $\theta_{\sigma \mu}$
   expressing   fluctuations 
 of the charge density and spin density for the $\mu$-band
 \cite{Suzumura_PTP}, where 
 the conjugate phase is introduced  by
$[\theta_{\nu \mu}(x),$ $\phi_{\nu' \mu'}(x')]_-$ $
= i \pi \delta_{\nu, \nu'} $ $\delta_{\mu, \mu'} $ ${\rm sgn}(x-x')$
 ($\nu,\nu'=\rho, \sigma$).
 Thus we obtain the total Hamiltonian  given by 
${\cal H}^d = {\cal H}_0 + {\cal H}_{I}$
  where ${\cal H}_0$  expresses bilinear terms of 
  density operators, and ${\cal H}_{I}$ denotes  nonlinear terms.
 Here we define   the new phase variables as 
  $Y_1 \equiv (X_1+X_2+X_3)/\sqrt{3}$, 
  $Y_2 \equiv (X_1-X_3)/\sqrt{2}$ and
  $Y_3 \equiv (-X_1+2X_2-X_3)/\sqrt{6}$ 
  where 
 $(Y_1,Y_2,Y_3) = 
(\theta_{\rho},\theta_{\rm C1},\theta_{\rm C2})$ and 
  $(X_1,X_2,X_3)
   =(\theta_{\rho +},\theta_{\rho 0},\theta_{\rho -})$ 
  and the same transformation is applied 
  to $\rho \rightarrow \sigma$ and also 
   to the conjugate 
  variables $\phi_{\nu \mu}$.
 Thus  ${\cal H}_0$ is rewritten as
${\cal H}_0 = 
  \sum_{\nu} (v_\nu/4\pi)  \int  dx
 [ K_\nu^{-1} \left(\partial \theta_{\nu} \right)^2
         +  K_\nu  \left(\partial \phi_{\nu} \right)^2]$
 with $\nu = \rho,{\rm C1},{\rm C2},\sigma,{\rm S1},{\rm S2}$,
 where 
$v_{\rho(\sigma)} = \vf [1+\!(-)U/\pi\vf]^{1/2}$,
$K_{\rho(\sigma)} =  [1+\!(-)Ua/\pi\vf]^{-1/2}$,
$v_{\rm C1} =v_{\rm S1} = v_{\rm C2} =v_{\rm S2} = \vf$ and
$K_{\rm C1}= K_{\rm S1} = K_{\rm C2}= K_{\rm S2} = 1$.
The Hamiltonian,   ${\cal H}_{I}$  is divided as 
  ${\cal H}_{I}= {\cal H}_{1}+ {\cal H}_{3} 
    + {\cal H}_{2}+{\cal H}_{\para}$  
 where  the respective term is written as ($i=1,3,2,\para$)
\begin{eqnarray}
{\cal H}_{i} &=& \sum_Z \frac{\vf}{\pi\alpha^2} \,\, G_{iZ} \int dx \,
  \cos \Theta_{iZ} \cdot h_{iZ} \virg 
\label{eq:H1}
\end{eqnarray}
 ($Z=A\simeq Z$)
 and $\alpha$ is 
 a  cutoff of the order of lattice constant and 
 $h_{iZ}$ denotes  the product  of the Majorana fermion operators
  introduced  to retain  the anticommutation relation of
  the field operators \cite{Schulz2}.
The phase variables of the backward scattering term
  with opposite spins, ${\cal H}_{1}$, are given by
  $\Theta_{1Z}=(2/\sqrt{3}) \theta_{\sigma} + \widetilde{\Theta}_{1Z}$
  where
  $\widetilde{\Theta}_{1A}= 
                \epsilon 
                \sqrt{2} \theta_{{\rm S1}} 
              - \sqrt{2/3} \, \theta_{{\rm S2}} $,
  $\widetilde{\Theta}_{1B}= 
                2\sqrt{2/3} \theta_{{\rm S2}}$,
  $\widetilde{\Theta}_{1C}= [ 
                (
                         \epsilon' 
                         \theta_{{\rm S1}} 
                       + \theta_{{\rm S2}} /\sqrt{3}
                )
              + \epsilon 
                (
                         \theta_{{\rm C1}} 
		       - \epsilon'
                         \sqrt{3}
			 (\theta_{{\rm C2}}+2\sqrt{6}\tperp x/\vf)
                )]/\sqrt{2}$,
  $\widetilde{\Theta}_{1D}= 
                \epsilon
                \sqrt{2} \theta_{{\rm C1}}
              - \sqrt{2/3} \theta_{{\rm S2}}$, 
  $\widetilde{\Theta}_{1E} = [
                (
                         \epsilon'
                         \theta_{{\rm S1}} 
                       + \theta_{{\rm S2}}  / \sqrt{3}
                )
              + \epsilon 
                (
                         \phi_{{\rm S1}} 
		       - \epsilon'
                         \sqrt{3}\phi_{{\rm S2}} 
                )]/\sqrt{2}$,
  $\widetilde{\Theta}_{1F} = 
                \epsilon
                \sqrt{2} \phi_{{\rm S1}} 
              - \sqrt{2/3} \theta_{{\rm S2}}$, 
  $\widetilde{\Theta}_{1G} = (
                \theta_{{\rm S2}} 
              + \epsilon' \sqrt{3} \theta_{{\rm C1}}
              + \epsilon   3 \phi_{{\rm C2}} 
              - \epsilon \epsilon' \sqrt{3} \phi_{{\rm S1}} )
          /\sqrt{6}$ and
  $\widetilde{\Theta}_{1H} = [
                (
                         \epsilon''
                         \theta_{{\rm S1}} 
                       - \theta_{{\rm S2}} /\sqrt{3}
                )
              + \epsilon' 
                (
                         \theta_{{\rm C1}} 
                       + \epsilon''
                         \sqrt{3}
			 (\theta_{{\rm C2}}+2\sqrt{6}\tperp x/\vf) 
                )
              + \epsilon 
                (
                         \epsilon''
                         3 \phi_{{\rm C1}} 
                       - \sqrt{3}\phi_{{\rm C2}} 
                )
              - \epsilon \epsilon' 
                (
                         \phi_{{\rm S1}} 
		       + \epsilon''
                         \sqrt{3} \phi_{{\rm S2}} 
                ) ]/2\sqrt{2}$.
In Eq.~(\ref{eq:H1}), the sum is taken implicitly
  with respect to 
 $\epsilon$, $\epsilon'$ and $\epsilon''$($=\pm$)
  which lead to the distinction for $h_{iZ}$ but not for $G_{iZ}$.
The umklapp scattering terms,
  ${\cal H}_{3}$,
  are obtained from ${\cal H}_{1}$ by replacing 
  $G_{1Z} \to G_{3Z}$ and
  $(\theta_{\sigma},\theta_{\rm S1},\theta_{\rm S2}) \leftrightarrow
   (\theta_{\rho},\theta_{\rm C1},
     \theta_{\rm C2}+2\sqrt{6}\tperp x/\vf)$. 
The phase variables of the forward scattering term
  with opposite spins, 
  ${\cal H}_{2}$, are expressed as 
  $\Theta_{2E}= [(
                         \theta_{{\rm C1}} 
                       - \epsilon'
                         \sqrt{3}
			 (\theta_{{\rm C2}}+2\sqrt{6}\tperp x/\vf )
                )
              + \epsilon 
                (
                         \phi_{{\rm S1}} 
		       - \epsilon'
                         \sqrt{3} \phi_{{\rm S2}} 
                )]/\sqrt{2}$,
  $\Theta_{2F}= \sqrt{2} \theta_{{\rm C1}}
              + \epsilon
                \sqrt{2} \phi_{{\rm S1}}$,
  $\Theta_{2G}= (\theta_{{\rm C1}}
              - \epsilon'
                \sqrt{3} \theta_{{\rm S2}} 
              - \epsilon \epsilon'
                \sqrt{3} \phi_{{\rm C2}} 
              + \epsilon
                \phi_{{\rm S1}} ) / \sqrt{2}$ and
  $\Theta_{2H}= [(
                         \theta_{{\rm C1}} 
                       + \epsilon''
                         \sqrt{3} 
                         (\theta_{{\rm C2}}+2\sqrt{6}\tperp x/\vf )
                )
              - \epsilon'
                (
                         \epsilon''
                         3        \theta_{{\rm S1}} 
                       - \sqrt{3} \theta_{{\rm S2}} 
                )
              - \epsilon \epsilon'
                (
                         \epsilon''
                         3       \phi_{{\rm C1}} 
                       - \sqrt{3}\phi_{{\rm C2}} 
                )
              + \epsilon
                (
                         \phi_{{\rm S1}} 
		       + \epsilon''
		         \sqrt{3} \phi_{{\rm S2}} 
                ) ] / 2\sqrt{2}$. 
 The forward scattering term with parallel spins, 
  ${\cal H}_{\para}$ is obtained from ${\cal H}_2$ by replacing 
  $G_{2Z} \to G_{\para Z}$($Z=E\sim H$) and
  $(\theta_{\rm S1},\theta_{\rm S2}) \leftrightarrow
   (\phi_{\rm S1},\phi_{\rm S2})$. 
Coupling constants are given by
$G_{iA} = G_{iB} = G_{iC} = G_{iD} = G_{iE} =
 G_{iF} = G_{iG} = G_{iH} = g_i/6\pi\vf$
($i=1,3$) and
$G_{iE} = G_{iF} = G_{iG} = G_{iH} = g_i/6\pi\vf$
($i=2,\para$)
where $g_\para \equiv g_{2^\para}-g_{1^\para}$.
Non-linear terms of forward scattering with the same $p$ branch
  are discarded because these effect is negligibly small. 
By assuming scaling invariance 
  with respect to $\alpha \to \alpha'=\alpha \e^{dl}$,
the second order RG equation for the interchain hopping is given by
\begin{figure}[t]
\begin{center}
\vspace*{6mm}
\leavevmode
\epsfysize=7.5cm
   \epsffile{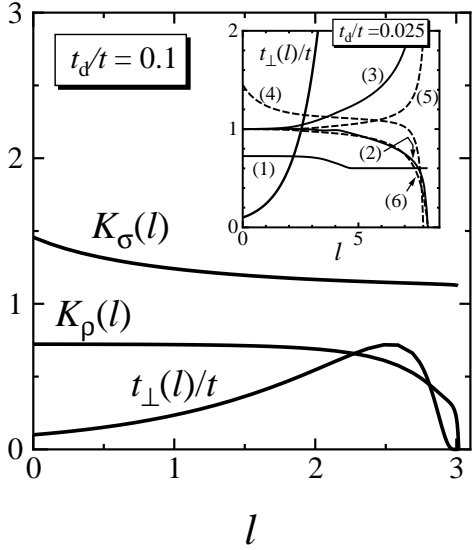}
\end{center}
\vspace*{-2mm}
\caption{
 Quantities 
$t_{\perp}(l)/t$, $K_{\rho}(l)$ and $K_{\sigma}(l)$ as a function of $l$
       for $U/t=4$, $t_\perp/t=0.1$ and $\td/t=0.1$.
The inset shows  $t_{\perp}(l)/t$, 
  $K_{\rho}$(1), $K_{\rm C1}$(2), $K_{\rm C2}$(3), 
  $K_{\sigma}$(4), $K_{\rm S1}$(5) and $K_{\rm S2}$(6) 
    for $U/t=4$, $t_\perp/t=0.1$ and $\td/t=0.025$.
}
\end{figure}
\begin{eqnarray}
\frac{d}{dl} \, \ttperp
   =  
\ttperp - F\left( \ttperp, \{G_{iZ}\} \right) K_{\rm C2},
\label{eq:interchain}
\end{eqnarray}
 where $\ttperp \equiv \tperp /(\vf\alpha^{-1})$,
$ F(\ttperp,\{G_{iZ}\})=
   [\Gbppzz^2 + \Gfpzzp^2 + \Gppzzp^2  ]
                     J_1(6\ttperp)
+    [\Gbpzpm^2 + \Gfpzpm^2 + \Gppzpm^2 ]
                    J_1(3\ttperp)
+    [ \Gupppp^2 + \Guppmm^2 + \Gupmmp^2 ]
                     J_1(4\ttperp)/3
+    [ \Guppzz^2 + \Gupzzp^2 + \Guzpzm^2 ]
                     J_1(2\ttperp)/3
+    \Guzzzz^2  J_1(8\ttperp)/3
+    \Gupzpm^2  J_1(\ttperp)/3$
and  $J_n$ is $n$-th Bessel function. 
 Equation (\ref{eq:interchain}) is solved together with 
 RG equations for $G_{iZ}$
where $l$ 
  is  related to energy scale $\omega$ or temperature $T$ by   
  $l=\ln(W/\omega)$ or $\ln(W/T)$ with $W(\equiv \vf \alpha^{-1})$ 
  being of the order of band width.  
 We take $\alpha=2a/\pi$ 
  \cite{Tsuchiizu_SCES}.
 It is  noted that 
 the r.h.s.~of eq.~(\ref{eq:interchain}) for small $\ttperp$
  is reduced to  
  $\ttperp[1-(G_1^2+G_2^2+G_{\para}^2+G_3^2)/2]$
  which becomes  the same as that of many chains
  \cite{Bourbonnais_U}.

In Fig.~1, the $l$-dependence of coupling constants is shown for 
  $\td/t=0.1$  and $\td/t=0.025$  
   with $U/t=4$ and $\tperp/t=0.1$.
 For $\td/t=0.025$ (inset),
   the interchain hopping $\tperp(l)$ becomes relevant corresponding to 
  deconfinement where 
   $K_\rho$  remains finite and 
  $K_{\rm C1}$, $K_\sigma$  and $K_{\rm S2}$  
    ($K_{\rm C2}$  and $K_{\rm S1}$)
   decrease to zero (become infinite).
 The  curves are calculated, for simplicity,  by setting  $J_n(l)=0$ 
   for  $l > l_n$  
  with $l_n$  corresponding the first node of the Bessel function, 
 although 
   such a treatment gives  negligible difference
    in the  numerical results.
 When  $K_\nu(l)$ decreases  to zero or 
   increases  to infinity,  
 the corresponding phase is locked leading to  
  a formation of gap,
 where relevant  coupling constants are  
  $G_{1B}(\to -\infty)$, $G_{1D}(\to -\infty)$,  
  $G_{1F}(\to -\infty)$, $G_{1G}(\to +\infty)$,
  $G_{2F}(\to -\infty)$ and $G_{2G}(\to +\infty)$.
 There are two kinds of gap for  charge fluctuations  and 
  three kinds of  
  gap for  spin fluctuations. 
For $\td/t=0.1$(main figure), one finds   confinement where
  $\tperp(l)$ decreases to zero after taking a  maximum.
The $l$-dependence of $K_\rho(l)$ implies  charge gap in 
  the total charge fluctuation, 
  which is in contrast to the case of deconfinement.
 For $\td/t=0.1$,   
  exponents $K_\nu(l)$ ($\nu=$C1, C2, $\sigma$, S1, S2) remain
  finite, i.e.,  $K_\nu(l)\simeq 1$  at $l\simeq 3$.
\begin{figure}[t]
\begin{center}
\vspace*{6mm}
\leavevmode
\epsfysize=7.5cm
   \epsffile{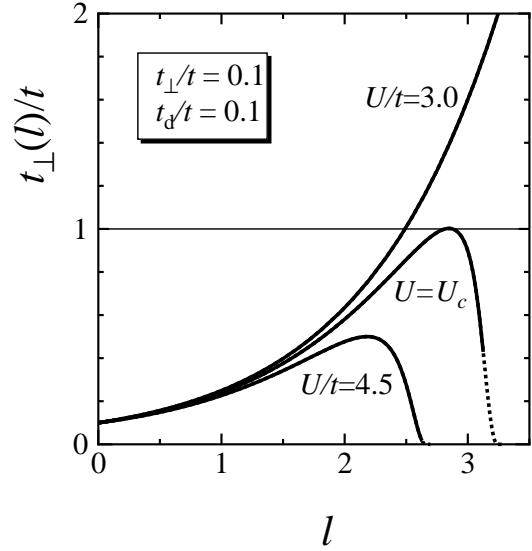}
\end{center}
\vspace*{-2mm}
\caption{
 Normalized quantity $t_{\perp}(l)/t$ 
   for  choices of $U/t = 3$, $U_c/t$ and 4.5 
  with $t_\perp/t=0.1$ and $\td/t=0.1$ where $U_c/t \simeq 3.7$.
The solid and dotted curves are explained in the text.
}
\end{figure}

In Fig.~2, $\tperp(l)/t$ is shown 
 with the fixed  $U/t= 3, U_c/t( \simeq 3.7)$  and 4.5   
 where   solid (dotted) curves denote $\tperp(l)/t$   for  
  $0<l<l_\Delta$ ($l>l_\Delta$) 
  with $K_\rho (l_\Delta) \equiv K_\rho/2$.
 The case for   $l>l_\Delta$ is invalid  
  since the magnitude of the umklapp scattering increases to infinity. 
 In the curve $\tperp(l)$, there is a  maximum given by  
   max$[\tperp(l)/t]\simeq 3.4$ (0.5) 
     at $l\simeq 4.0$ (2.2) 
     for $U/t=3.0$ (4.5) although    the maximum  for $U/t=3.0$
       is located  in the region of $l>l_\Delta$. 
Thus the case for $U/t=3.0$ (4.5) corresponds 
   to deconfinement (confinement).
 The  boundary between  
    confinement and deconfinement is determined 
  by the condition that max$[\tperp(l)/t]\simeq 1$ at $U=U_c$.

\begin{figure}[t]
\begin{center}
\vspace*{6mm}
\leavevmode
\epsfysize=7.5cm
   \epsffile{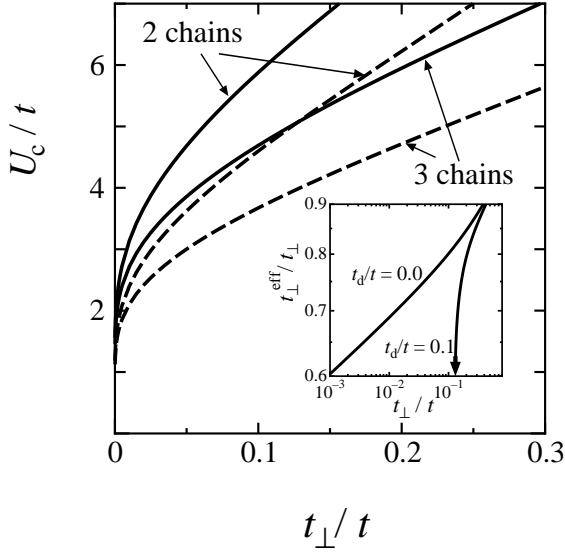}
\end{center}
\vspace*{-2mm}
\caption{
The critical values  $U_c$ as a function of $\tperp$ 
  with fixed $\td/t=0.05$(solid curves) and 0.1(dashed curves)
  for three-coupled chains  and two-coupled chains
 \cite{Tsuchiizu_SCES}.
 The inset denotes 
  the log $\tperp^{\rm eff}/\tperp$-log $\tperp$ plot 
  for $\td/t=0.0$ and 0.1 with fixed $U/t=4.0$  where 
 the arrow shows the critical value for the  confinement. 
}
\end{figure}
In Fig.~3, the $\tperp$-dependence of $U_c$ is shown 
  for $\td/t=0.05$(solid curves) and 0.1(dashed curves)
   where confinement (deconfinement) is 
  obtained for $U>U_c$ ($U<U_c$).
For comparison, the corresponding results for two-coupled chains
  \cite{Tsuchiizu_SCES} are also shown.
The critical values for three-coupled chains
  is smaller than that for two-coupled chains.
 Since the RG equation 
   of two-coupled chains corresponding to Eq.~(\ref{eq:interchain})
     includes  the Bessel function with only  $J_1(8\ttperp)$,
 the effect of umklapp scattering for three-coupled chains
  is stronger than that for two-coupled chains.
  The effective  interchain hopping $\tperp^{\rm eff}$ 
   can be evaluated  from 
  $\tperp^{\rm eff}=t\exp[-l_{\rm eff}]$ 
  where $\tperp (l_{\rm eff})/t=1$
  \cite{Tsuchiizu_PRG}. 
  In the inset, 
  $\tperp^{\rm eff}/\tperp$  is shown 
  as a function of $\tperp$ on a logarithmic scale.
 The power-law behavior of  
    $\tperp^{\rm eff,0}$ $(\equiv \tperp^{\rm eff} (g_3=0)$)
   for small $\tperp$ 
   is consistent with the analytical formula
  $\tperp^{\rm eff,0} \simeq \tperp (\tperp/W)^{\alpha_0/(1-\alpha_0)}$
  with $\alpha_0=(K_\rho+K_\rho^{-1}+K_\sigma+K_\sigma^{-1}-4)/4$
  \cite{Bourbonnais_U}.
In the presence of the dimerization, $\tperp^{\rm eff}$ is reduced 
 from that of 
   the power-law behavior and has a critical value below 
  which the confinement occurs.

\begin{figure}[t]
\begin{center}
\vspace*{6mm}
\leavevmode
\epsfysize=7.5cm
   \epsffile{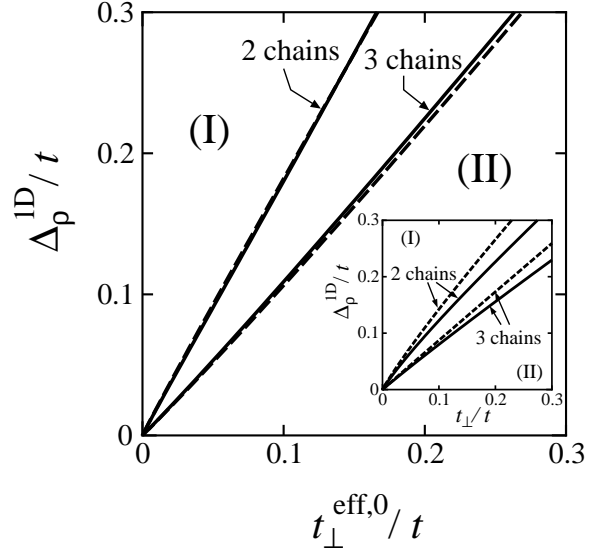}
\end{center}
\vspace*{-2mm}
\caption{
The phase diagram of confinement (I) and deconfinement (II)
  in case of $\td/t=0.05$ (solid line) and 0.1 (dashed line)
  on a plane of the effective interchain hopping $t_\perp^{\rm eff,0}$
   and $\Delta_\rho^{\rm 1D}$.
 The inset denotes the corresponding phase diagram on the plane of 
  $\tperp$ and $\Delta_\rho^{\rm 1D}$.
}
\end{figure}
 The confinement-deconfinement transition is examined in terms of 
   the charge gap induced by the umklapp scattering.
 The charge gap for the single chain 
   is  obtained   as the  function of $U$ and $\td$ by a method 
 $\Delta_\rho^{\rm 1D}=W \exp[-l_\Delta]$
  with $K_\rho(l_\Delta)=K_\rho/2$ for  $\tperp=0$.
 Such   gap has a meaning of a characteristic energy 
  of the umklapp scattering
    even for  the deconfined  region  
     in which the charge gap  is reduced to zero 
         due to the presence of 
         the misfit for all the nonlinear terms of umklapp scattering.
In the inset of Fig.~4, 
  a phase diagram of confinement (I)  and deconfinement (II)
  is shown on the plane of the bare interchain hopping $\tperp$ 
  and $\Delta_\rho^{\rm 1D}$ where the boundary for two-coupled chains
  \cite{Tsuchiizu_SCES} are also shown for comparison.  
Although the ratio of $\Delta_{\rho}^{\rm 1D}/\tperp$ 
  is nearly constant, 
  the curve is slightly convex upward for small $\tperp$.
In the main figure of Fig.~4, the phase diagram with the same parameter 
 is shown on the plane
  of $\tperp^{\rm eff,0}/t$ and $\Delta_{\rho}^{\rm 1D}/t$. 
  The quantity  $\tperp^{\rm eff,0}$ denotes the 
  effective interchain hopping, which is 
  renormalized by the intrachain interaction without umklapp scattering.
The ratio of $\Delta_{\rho}^{\rm 1D}$ to $\tperp^{\rm eff,0}$ 
 at the boundary is estimated as follows when  $0.05<\tperp/t<0.3$.
 The ratio for three-coupled chains is given by 
  $1.0 \lsim \Delta_{\rho}^{\rm 1D}/\tperp^{\rm eff,0} \lsim 1.1$
  while that for two-coupled chains is given by 
  $1.8 \lsim \Delta_{\rho}^{\rm 1D}/\tperp^{\rm eff,0} \lsim 1.9$.
By noting that the $\td$-dependence of the boundary is very small,
 it turns out that  
 the confinement-deconfinement transition is determined by
  the competition between the charge gap and the effective 
   interchain hopping energy.

 We briefly discuss the case for OBC where 
 the RG equation for the interchain hopping takes more complicated form
   due to twelve coupling constants for umklapp scattering. 
 The small  difference between PBC and OBC is expected
   since   the RG equation for the interchain hopping
 has the same limiting form as that of PBC for small $\ttperp$.
Actually,  we find 
  almost the  same boundary as Fig.~3 
 when  
 the solution of single 
    chain is substituted for  the RG equation of interchain hopping.

In summary,  
   the confinement-deconfinement 
  transition 
  in the three-coupled Hubbard chains with dimerization 
  has been shown for PBC 
 when the effective  interchain hopping energy becomes of the order   
   the charge gap induced by the umklapp scattering.

The authors thank H.~Yoshioka for useful discussions.
This work was supported by a Grant-in-Aid for Scientific 
  Research from the Ministry of Education, Science, Sports and
  Culture (Grant No.09640429), Japan.

\end{document}